\documentstyle[12pt,aaspp4]{article}

\lefthead{Bekki Kenji}
\righthead{Formation of nuclear gaseous disk}

\begin{document}
\title{Gas fueling and  nuclear gaseous disk formation
in merging between a central massive black hole and
a gas clump}

\author{Kenji Bekki} 
\affil{Division of Theoretical Astrophysics,
National Astronomical Observatory, Mitaka, Tokyo, 181-8588, Japan} 

\begin{abstract}

We numerically investigate dynamical evolution
of a merger between a  central massive black hole (MBH)
and a gas clump with the mass of
$10^6$ $-$ $10^7$ $M_{\odot}$   in the central tens pc of a galactic bulge.
We found that strong tidal gravitational field of the MBH
transforms the initial spherical clump into
a moderately thick gaseous disk (or torus) around the MBH.
The developed disk is also found to show rotation, essentially because 
the tidal field  changes  efficiently  the orbital angular momentum
of the clump  into intrinsic angular momentum  of the disk.  
Furthermore about a few percent of gas mass 
(corresponding to a few $10^5$ $M_{\odot}$) in the clump
is found to be transferred to the central
sub-parsec region around the MBH
within an order of $10^6$ yr. 
We thus suggest that successive merging of  gas clumps
onto a MBH
can not only be  associated closely with the formation
of nuclear disk around the MBH 
but also can  provide  gas fuel for the  MBH.

\end{abstract}


\section{Introduction}
It is one of longstanding and remarkable problems on
formation and evolution of active galactic nuclei (AGN)
how interstellar gas can be fueled to the central
massive black hole (hereafter referred to as MBH) in galaxies 
(e.g., Shlosman, Frank, \& Begelman 1989; Shlosman, Begelman, \& Frank 1990).
The central core of this fueling problem
is on how the angular momentum of interstellar gas can be 
reduced by more than several orders of magnitude in order that
the gas can be transferred into the central sub-parsec of a galaxy
(Shlosman et al. 1990).
Shlosman et al. (1989) proposed the so-called `bars within bars
scheme' as a possible mechanism for fueling AGN. They showed
analytically that for suitable conditions the gas inflow induced by
bar instability again produces the self-gravitating disk in a smaller region
and a new small bar develops, which induces further gas inflow.
Friedli \& Martinet (1993) numerically studied the dynamical evolution
of doubly nested bars and found that for suitable conditions the gas
inflow to the nuclear region is derived by the dissolution of the secondary
smaller bar in a two-bar system.
Bekki \& Noguchi (1994) found that 
the dynamical heating by two sinking cores and subsequent dissipative 
cloud-cloud collisions in  a merger can drive a larger fraction of gas to
the central 10 pc.
These theoretical studies essentially stressed the importance
of non-axisymmetric and time-dependent gravitational  potential
in gaseous inflow into the central sub-parsec regions.

Shlosman \& Noguchi (1993) presented an alternative and important
point of view that for a strongly self-gravitating disk,
gas clumps formed from local gravitational instability 
can quickly fall toward the nuclear region around the MBH
owing to dynamical
friction.  
Noguchi (1998) furthermore demonstrated that  
these massive clumps are more likely to be formed
in the early disk formation phase when disk galaxies
have a larger amount of gas.
These two numerical studies stressed the importance of 
discrete gas clumps (not continuous and diffuse gas)
in gas fueling processes.
Recent observational studies on ultra-luminous infrared
galaxies (ULIRGs), 
some of which are suggested to contain MBHs (Sanders et al. 1988),
have revealed that 
most of ULIRGs show compact blue knots  
that are probably very young star clutters with 
the masses of $10^{5}-10^{9}$ $M_{\odot}$ (e.g., Surace et al 1998).
Shaya et al. (1994) furthermore 
revealed  a number of bright and possibly young
star clusters in the core of ULIRG Arp 220 and suggested that these
clusters can be very quickly transferred to the inner tens pc
within an order
of $10^8$ years owing to dynamical friction.
Since these observed clusters
could be formed by massive gaseous clumps (e.g., massive molecular
clouds), 
the above  observational results lead us to
suggest that gaseous clumps
are also transferred to the vicinity of MBHs.

The purpose of this Letter 
is to investigate numerically  merging between 
a central massive black
hole with the mass of 
$\sim$ $10^7$ $M_{\odot}$ and a spherical gaseous clump 
with the mass of $\sim$ $10^7$ $M_{\odot}$ and the size of
$\sim$ 10  pc.
We  demonstrate  that strong tidal field of the MBH
can transform the clump into a moderately 
thick gaseous disk (or torus):
The MBH can form the surrounding nuclear disk for itself. 
We suggest that the viscosity of the gaseous disk
could transfer the gas furthermore to the  vicinity of
the MBH and grow the MBH. 
Based on the present numerical results, 
we furthermore provide several implications on recent observational results,
such as 
active  star formation around MBH in the Galaxy 
(e.g., Morris \& Serabyn 1996),
apparently double M31 nuclei (e.g., Kormendy \& Bender 1999),
and nuclear gaseous disks observed
in some galaxies such as NGC 4261 (e.g., Jaffe et al. 1993). 
Importance of tidal disruption of stars passing by
a central MBH in the accretion disk was already suggested
by Gurzadyan \& Ozernoy (1979) and there are several numerical studies 
investigating tidal disruption of $star$  clusters 
(e.g., Charlton \& Laguna 1995; Emsellem \& Combes 1997; Bekki 2000).
However, it is highly uncertain how a merger between a MBH
and a massive $gas$ clump dynamically evolves. 
Thus we consider that the present study not only 
presents a solution of the fueling problem of AGN
but also provides a clue to the origin of nuclear structure
of galaxies.

\section{Model}

We consider a 
merger between a central massive black hole
(MBH) with the mass of $M_{\rm BH}$
and a gaseous clump  with
the mass of $M_{\rm gas}$ and the size of
$R_{\rm t}$ in the central region
of a galactic bulge.
From now on, all the mass and length are measured in units of
$M_{\rm BH}$ and  $R_{\rm t}$, respectively, unless specified. 
Velocity and time are
 measured in units of $v$ = $ (GM_{\rm BH}/R_{\rm t})^{1/2}$ and
$t_{\rm dyn}$ = $(R_{\rm t}^{3}/GM_{\rm BH})^{1/2}$, respectively,
where $G$ is the gravitational constant and assumed to be 1.0
 in the present study.
If we adopt $M_{\rm BH}$ =  $10^{7}$ ($10^{8}$) $M_{\odot}$ and
$R_{\rm t}$ = 10  pc as a fiducial value, then $v$ = 6.5 $\times$
10 (2.1 $\times$ $10^2$) km/s  and  $t_{\rm dyn}$ = 1.49 $\times$ $10^{5}$ 
(4.72 $\times$ $10^4$)  yr,
respectively.
Both the MBH and the clump are assumed to feel
external gravitational field of the bulge component.
The MBH is assumed to be initially located at the center
of the bulge and the initial separation between the MBH
and the clump is set to be 3$R_{\rm t}$.
For the radial density profile of the bulge, 
we adopt the universal profile proposed by Navarro, Frenk,
\& White (1996). 
We assume that the scale length 
is equal to
10$R_{\rm t}$ and determine the central density
so that total mass of the bulge within 200$R_{\rm t}$ ($\sim$ 2 kpc)
is 200$M_{\rm BH}$. 
The adopted ratio of bulge mass to MBH one is well within
a reasonable value derived by Faber et al. (1997). 
The total mass of the bulge within $R$ = 1.0
(the central 10 pc), where $R$ is the distance from the center of
the bulge,
is hereafter represented by
$M_{\rm gal}$.
We model a $n$ = 1
polytropic gas sphere in hydrostatic equilibrium
for the initial density profile of the clump.
For the fiducial model, the mass ratio  $M_{\rm gas}$/$M_{\rm BH}$
and $M_{\rm gal}$/$M_{\rm BH}$ are set to be 1.0 and 0.3, respectively.
For this case, the tidal radius of the MBH is nearly the same
as the size of the clump.

We describe the gaseous nature of the clump
 by using the Smoothed Particle
Hydrodynamics method 
(introduced by Lucy 1977 and Gingold \& Monaghan 1977)
in which the fluid is modeled as a collection of fluid elements.
The gas clump is assumed to be polytropic gas with
the equation of state represented by $P$ = $K {\rho}^{\gamma}$,
where $P$, $K$, $\rho$, and $\gamma$ are
pressure, constant, density, and the ratio of specific
heat (2 for the adopted value of $n$ =1 polytrope),
respectively.
The initial orbital plane of the merger
is  assumed to be exactly the same as the $x$-$y$ plane.
Initial $x$ and $y$ position ($x$, $y$) is set to be
(0,0) for the MBH and (3,0) for the clump in all models.
Initial $x$ and $y$ velocity ($V_{\rm x}$, $V_{\rm y}$)
is set to be (0,0) for the MBH and 
(0, 0.75$V_{\rm cir}$) for the clump 
in the fiducial model, where $V_{\rm cir}$ is the circular velocity
at the initial position of the clump.
The  number of particles 
for  the clump is 5000 and  
the parameter of gravitational softening is set to be fixed at
2.97 $\times$ $10^{-2}$ in our units 
for all the simulations. 
By using the TREESPH code originally proposed by
Hernquist \& Katz (1989), we carry out all the calculations
related to the above dynamical evolution of mergers.
Using the above model, we mainly describe  morphological,
structural,  and kinematical
properties of the merger remnant in the fiducial model.
Moreover we summarize briefly the dependence of mass distribution
of merger remnant on the initial merger parameters such as $M_{\rm BH}$,
$V_{\rm x}$, and $V_{\rm y}$.
More details on the parameter dependences will be described in our
future papers.

\placefigure{fig-1}
\placefigure{fig-2}
\placefigure{fig-3}

\section{Result}

Figure 1 shows the time evolution of mass distribution 
of a gaseous clump in the fiducial  merger model.
As the clump becomes close to the MBH,
strong tidal gravitational field draws out
the clump and consequently the outer part of the clump
winds itself round the MBH ($T$ = 6).
During this  strong tidal disruption,
angular momentum redistribution of the merger proceeds
very efficiently and consequently
a gaseous tidal tail is formed ($T$ = 10 and 14).
The MBH's tidal field finally transforms the initially
spherical shape of the clump into
a moderately thick gaseous disk with a remarkable tidal tail
after merging ($T$ = 16 and 20).
During this morphological transformation,
the initial orbital angular momentum of the clump 
is efficiently changed into the intrinsic angular momentum
of the forming disk around the MBH.
As a natural result of
this, the developed disk clearly shows rotation. 
The vertical scale height of the developed disk
depends mainly on the initial half-mass radius
of the clump (for the present
polytropic gas model).

As is shown in Figure 2, 
the developed nuclear gaseous disk
clearly shows  strong asymmetry
both in surface density profile
and in rotation curve along $x$ and $y$ axis
owing to the presence of one tidal arm.
The radial density profile
has a flat peak around $-0.4$ $<$ $x$ ($y$) $<$ 0,
where some fraction of gas 
is accumulated in the  vicinity of the MBH.
Owing to the self-gravity of the developed disk
and the bulge's gravitational field (and the asymmetric
mass distribution),
the disk does not so clearly show the Keplerian rotation
profile, though the gravitational field of the point mass
is rather strong. 
As merging proceeds, the gas of the clump can be efficiently
accumulated around MBH (See Figure 3).
The gas mass within $R$ = 0.1 in our units (corresponding
to 1 pc),
where $R$ is the distance from the MBH,
rapidly increases at  $T$ = 8 and finally amounts to $\sim$ 0.04   
in our units 
(corresponding to 4 $\times$ $10^5$ $M_{\odot}$)
at $T$ = 20.
This result implies that merging between a MBH and a gas clump
can very efficiently transfer gas to the surrounding
of the MBH and consequently
form an AGN: A MBH can provide gaseous fuel for it for itself
because of its strong tidal gravitational field.

It depends mainly on the mass ratio of $M_{\rm gas}/M_{\rm BH}$ 
whether a nuclear disk around MBH is formed after merging 
(and tidal interaction).
As is shown in  Figure 4, the gas clump is only slightly
distorted by the tidal gravitational field of the MBH
for the model with larger $M_{\rm gas}/M_{\rm BH}$
(i.e., smaller $M_{\rm BH}$ and weaker tidal field of MBH) so that
no disks can be formed.
For the model with smaller $M_{\rm gas}/M_{\rm BH}$
(i.e., larger $M_{\rm BH}$),
on the other hand, the spherical shape of the clump
is finally transformed into a moderately thick torus
(or a disk with a central hole) around the MBH.
Furthermore the morphology of the central gaseous structure 
depends on the present model parameters.
In the model with rather small initial angular momentum
($V_{\rm x}$ = 0.0 and $V_{\rm y}$ = 0.2$V_{\rm cir}$),
the clump merges with the MBH to form a spherical clump
with the  MBH gravitationally trapped in the  
central region of the clump (i.e., no tidal disruption
of the clump).
A larger eccentric gaseous disk with the steeper radial
density profile is formed in the model
with larger initial orbital angular momentum 
($V_{\rm x}$ = $-0.5V_{\rm cir}$ and  $V_{\rm y}$ = 0.75$V_{\rm cir}$).
Only a considerably  small gaseous disk around the  MBH 
is formed owing to gas transfer in the model
in which the MBH and the clump 
encounter in a hyperbolic way ($V_{\rm x}$ =  $-2.0V_{\rm cir}$ 
and  $V_{\rm y}$ = 0.75$V_{\rm cir}$).
These dependences imply that 
 nuclear gaseous structure around a MBH in  a galaxy
is determined by orbital parameters of a merger
and the mass of the MBH.

\placefigure{fig-4}

\section{Discussion}

The present numerical results provide the following
five implications on the nuclear structure and
activities of galaxies.
Recently Hubble Space Telescope ($HST$) observations 
have revealed nuclear gaseous disks with the size
ranging from $\sim$ 
 1 pc to $\sim$ 1 kpc in nearby galaxies (e.g., Jaffe et al. 1993, 1999).
Our results firstly suggest that
not the continuous gaseous inflow but the
successive and sporadic merging of discrete 
gas clumps (or gas-rich small dwarf galaxies) onto a MBH can be  one of 
essentially important and basic processes
of the nuclear gaseous disk formation.
Secondly, our results provide a more detailed 
description of gas fueling from the central 10 pc to 
sub-parsec around MBHs.
After the formation of a nuclear disk with the size of $\sim$ 10 pc
due to merging between a MBH and a gas clump,
the gas in the disk can be transferred to the vicinity of the MBH
within the viscosity time scale of $10^6$ yr  (e.g., Jaffe et al. 1999). 
Even if a MBH is not so large as to tidally disrupt
a clump (for the present model with smaller  $M_{\rm BH}$), 
the MBH trapped within the central
high density core of the clump could grow owing
to Bondi-type (Bondi 1952) accretion.  
We accordingly point out that the merging
could provide ideal gas fueling 
for MBHs.
We furthermore suggest that since gas clumps are more likely to 
be formed and transferred
to the nuclear regions in  higher  redshift galaxies (Noguchi 1998),
MBHs can more rapidly grow  owing to gaseous accretion
and consequently show more pronounced nuclear activities
 at higher redshift.

Thirdly, the present model  (with larger $M_{\rm BH}$) showing  
the formation of gas torus implies that
the obscuring torus 
proposed for explaining the origin of the dichotomy between
Seyfert 1 and 2 galaxies
in the unified model (e.g., Antonucci 1993)
could be associated with merging between a MBH and gas clumps. 
Since gas rich clumps are more likely to be formed
in gas-rich disks  with the gas fraction
larger than 0.1 (Shlosman \& Noguchi 1993),
the present study
suggests  that late-type disk galaxies are more likely to
show Seyfert 2 type activities with possible obscuring torus.
Fourthly, the present model implies that
if star formation is included and assumed to be dependent
on gaseous density, high density regions of the developed
nuclear disk around MBH
can efficiently form stars: `Mini' starburst is triggered by
merging between MBHs and gas clumps.
We furthermore suggest that a young central cluster of blue stars
observed in the very center of the  Galaxy (e.g., Morris \& Serabyn 1996) 
is a fossil record of  the past merging event.  
Fifthly, the present study provides a new clue to the origin
of M31 nucleus having  two distinct brightness
peaks with the separation of $\sim$ 2 pc (See 
Kormendy \& Bender 1999 for recent high-resolution observational
results). 
Although Tremaine (1995) recently proposed a new idea that
the M31's nucleus is  actually a single thick eccentric disk
(with the mass of the order of $10^6$ $M_{\odot}$) 
surrounding the central MBH,
it is highly unclear how the disk was formed.
Our results imply that if a  smaller gaseous clump
with the mass of $\sim$ $10^6$
merges with the MBH and if the later star formation
transforms the formed gaseous disk into
the stellar one, the proposed M31's eccentric disk is formed
(In this point, Bekki 2000 has already presented an alternative idea
on the M31's nucleus formation).
Although 
some of the above several suggestions could be somehow over-interpretation
of the present numerical results,
we lastly stress that merging between a MBH and gas clumps 
can be  an important driver of nucleus evolution in galaxies.

\clearpage


\figcaption{
Mass distribution of a merger between
the central MBH and a gaseous clump   projected
onto $x$-$y$ plane (upper six panels) and onto $x$-$z$ one
(lower six ones) at each time $T$
for the fiducial model 
with $M_{\rm gas}/M_{\rm BH}$ = 1.0,  $M_{\rm gal}/M_{\rm BH}$
=0.3, $V_{\rm x}$ = 0, and $V_{\rm y}$ = 0.75$V_{\rm cir}$. 
$T$ (in our units)  is indicated in the upper
left-hand corner of each panel.
The MBH is initially located exactly in the center of 
a bulge component (i.e., the $x$ and $y$ position ($x$,$y$) is (0,0)).
Here the scale is given in our units (corresponding to 10 pc)
and each of the 12 frames measures 80 pc (8.5 length units)
on a side.   
Note that strong tidal field of the MBH transforms the
initial spherical shape of the clump into a moderately thick 
nuclear gaseous disk. 
\label{fig-1}}

\figcaption{
The radial density profile (upper) and  the rotation curve (lower)
along $x$-axis (solid line)
and $y$-axis (dotted one) at $T$ = 20.0 for the developed nuclear disk.
Here the scale for length and velocity is given in our units
and the surface density scale is given in units of 100 $\times$ 
$M_{\rm BH}$/${R_{\rm t}}^2$ for clarity. 
Note that since the initial orbital angular momentum of the merger
is efficiently changed into the intrinsic
angular momentum of the disk,
the disk  shows clearly rotation.
Note also that structure and kinematics of the disk clearly show
asymmetry in the radial distributions
owing to the one tidal arm formed during merging.
\label{fig-2}}

\figcaption{
Time evolution of 
gas mass accumulated within $R$ = 0.1 (solid) and 0.5 (dotted),
where $R$ is the distance from the MBH.
The accumulated mass is 0.14 
for $R$ $<$ 0.5 and 0.04 for $R$ $<$ 0.1.
The gas fueling rate in this model for 10 $\le$ $T$ $\le$ 20 
is estimated to be $\sim$ 2.7 $M_{\odot}$ ${\rm yr}^{-1}$ 
for $R$ $<$ 0.1 (the central 1 pc). 
Thus this figure indicates that merging between a MBH
and a gas clump can be a possible candidate of ideal
fueling mechanisms.
\label{fig-3}}

\figcaption{
Final mass distribution projected
onto $x$-$y$ plane (upper two)
and onto $x$-$z$ one (lower two) for the model with
$M_{\rm BH}$ = 0.1, $M_{\rm gas}/M_{\rm BH}$ = 10.0,
and $M_{\rm gal}/M_{\rm BH}$ = 3.0 (left panel, represented by `Smaller MBH')
and for the model with
$M_{\rm BH}$ = 10.0, $M_{\rm gas}/M_{\rm BH}$ = 0.1,
and $M_{\rm gal}/M_{\rm BH}$ = 0.3 (right panel, represented by `Larger MBH')
Here parameter values other than those specified above
are exactly the same as those of the fiducial model
for each of the two models.
Note that only for the larger MBH model, a moderately thick
gaseous torus (or a disk with the central hole) is formed around MBH.
\label{fig-4}}

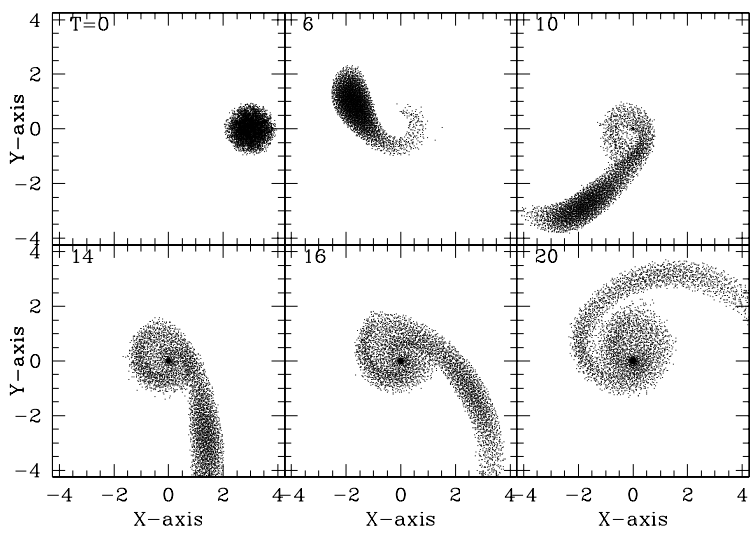
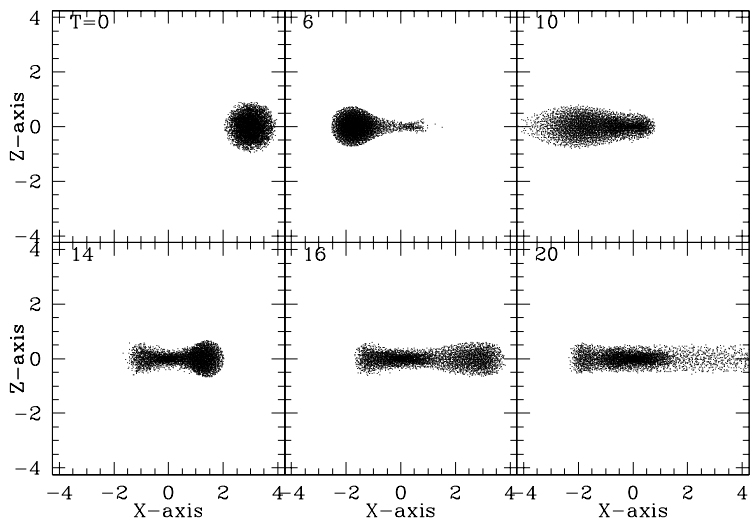
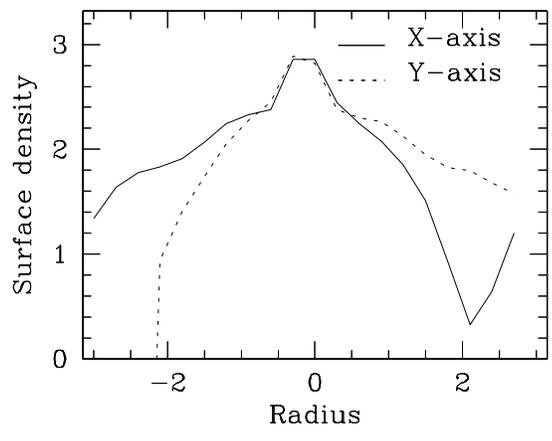
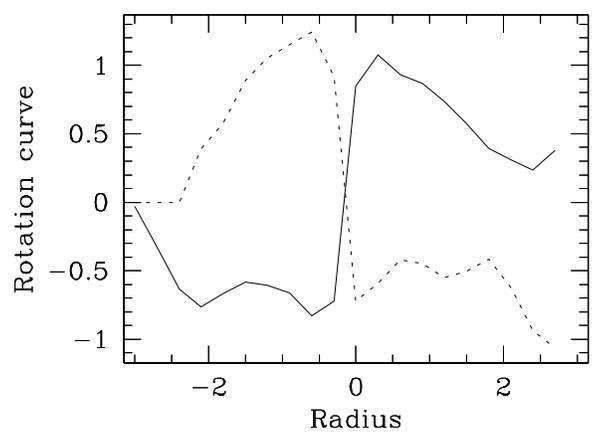
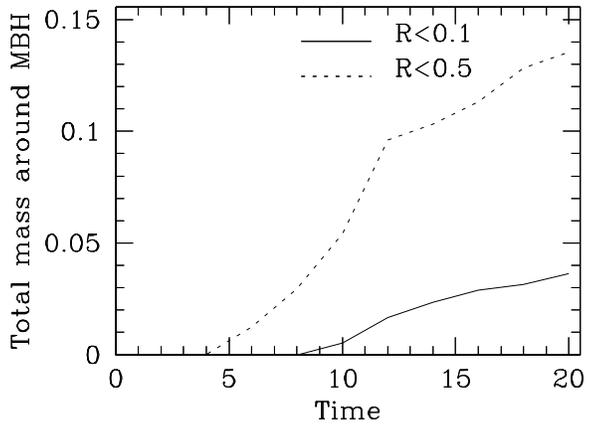
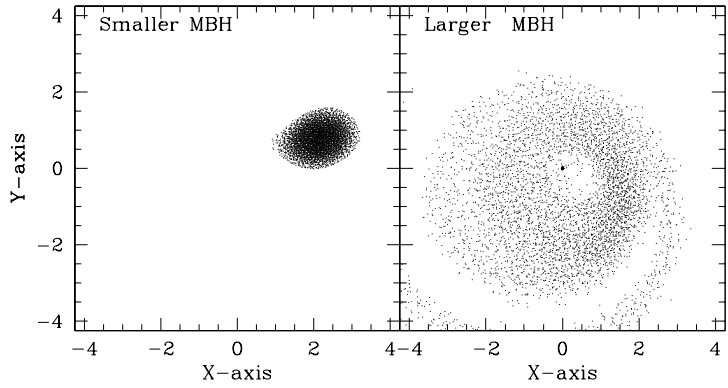
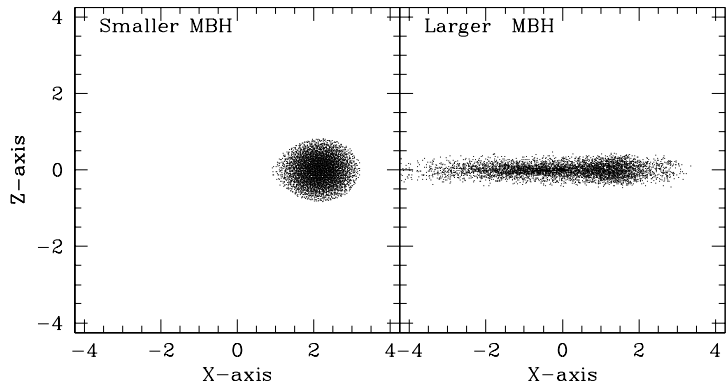
\end{document}